\documentstyle[12pt]{article}
\newcommand{\beq}{\begin{eqnarray}}
\newcommand{\eeq}{  \end{eqnarray}}

\newcommand{\dv}[2]{\frac{\textstyle #1}{\textstyle #2}}

\def \gapx {\lower 2pt \hbox{ $\buildrel>\over{\scriptstyle{\sim}}$ }}
\def \lapx {\lower 2pt \hbox{ $\buildrel<\over{\scriptstyle{\sim}}$ }}
\def\prd#1 { Phys.\ Rev.\ D {\bf #1 }}
\def\prc#1 { Phys.\ Rev.\ C {\bf #1 }}
\def\npb#1 { Nucl.\ Phys.\ B {\bf #1 }}
\def\plb#1 { Phys.\ Lett.\ B {\bf #1 }}
\def\zpc#1 { Z.\ Phys.\ C {\bf #1 }}
\def\prl#1 { Phys.\ Rev.\ Lett.\ {\bf#1 }}

\newcommand{\ra}{\rightarrow}

\newcommand{\vv}[1]{ {\bf #1}}
\def\lp{\lambda^{\prime} }

\def\an{ {J,J_3} }
\def\ji{ dx\ d^2{\bf k}_\bot}

\def\et{et al.}

\begin{document}
\baselineskip=5mm
\newcount\sectionnumber
\sectionnumber=0
\pagestyle{empty}
\begin{flushright}{ UTPT--95--2 }
\end{flushright}

\vspace{8mm}
\begin{center}
{\bf {\huge Weak Decays in the light--front Quark Model} }\\
\vspace{6mm}
Patrick J. O'Donnell, Q.P. Xu  \\
Physics Department,\\
University of Toronto,\\
Toronto, Ontario M5S 1A7, Canada,\\
\ \\
and \\
\ \\
Humphrey K.K. Tung\\
Department of Applied Physics\\
Hong Kong Polytechnic University\\
Hung Hom, Hong Kong.
\end{center}

\vskip 20pt
\centerline{\bf Abstract}

We   study   the   form   factors   of    heavy--to--heavy    and
heavy--to--light weak decays using the light--front  relativistic
quark  model.  For  the   heavy--to--heavy   $B  \ra  D^{(\ast)}$
semileptonic  decays we calculate the  corresponding  Isgur--Wise
function    for   the   whole    kinematic    region.   For   the
heavy--to--light  $B\ra P$ and $B\ra V$  semileptonic  decays  we
calculate the form factors at $q^2 = 0$;  in particular,  we have
derived the dependence of the form factors on the $b$--quark mass
in the  $m_b  \ra  \infty$  limit.  This  dependence  can  not be
produced  by  extrapolating  the  scaling  behavior  of the  form
factors at $q^2_{max}$  using the single--pole  assumption.  This
shows that the $q^2$  dependence  of the form  factors in regions
far away  from the  zero--recoil  could be much more  complicated
than that predicted by the single--pole assumption.

\newpage
\pagestyle{plain}
\section{Introduction}

In  the  last  few  years,  great   progress  has  been  made  in
understanding  weak decays of hadrons  containing  heavy  quarks.
The heavy quark symmetry, which appears in the heavy quark limit,
can  simplify  many  aspects of the weak decays of heavy  hadrons
\cite{heavy}.  Due to the heavy quark  symmetry all form  factors
in the heavy--to--heavy type of decays such as
$B \ra D^{(\ast)} e {\bar \nu_e}$
($D^{(\ast)}=D$ or $D^\ast$)
can be related, in the heavy quark  limit, to a single  universal
function   called  the  Isgur--Wise   function.  The  Isgur--Wise
function  is of  nonperturbative  origin  and has  been of  great
interest  to  both  theoretical  and  experimental   studies.  In
particular, the  Isgur--Wise  function of the $B \ra  D^{(\ast)}$
semileptonic      decays     has     been     widely      studied
\cite{rela,lattice,sumrule,isgw,old}.

The  heavy  quark  symmetry  can  also  shed  some  light  on the
heavy--to--light  type of  weak  decays.  For  example,  one  can
derive the  dependence  of form  factors  (there is more than one
form  factor in this  case) on the heavy  quark  mass in the zero
recoil  region, i.e near  $q^2_{max}$  \cite{iw}.  However,  away
from the zero--recoil  region, one still needs a model--dependent
method to understand the form factors.

In this paper we study the form factors of  heavy--to--heavy  and
heavy--to--light weak decays using the light--front  relativistic
quark  model.  The  light--front  relativistic  quark  model  was
developed  quite  a long  time  ago  and  there  have  been  many
successful applications \cite{model,jaus,ox1,ox2}.  Here we use
this  model  to  calculate  the  Isgur--Wise   function  for  the
heavy--to--heavy $B \ra D^{(\ast)}$  semileptonic  decays.  It is
known that the light  front  model  usually can only work at $q^2
\le 0$.  However, it is possible to use the results at $q^2=0$ to
get the  Isgur--Wise  function  for the  whole  kinematic  region
\cite{nr}.

We also study the form  factors  in the  heavy--to--light  decays
such as $B \ra \pi$ and $B \ra \rho$  semileptonic  decays in the
heavy $b$--quark  limit.  In particular, we are interested in the
dependence  of these form  factors on the  $b$--quark  mass $m_b$
since the  pole--dominance  assumption  for the form  factors  is
usually   used  to  go  away   from  the   zero--recoil   region.
Heavy--to--light  weak  decays are  especially  sensitive  to the
$q^2$  dependence  of the form factors  with such an  assumption.
The scaling  behavior of these form factors at $q^2=0$  allows us
to compare with the pole--dominance  assumption and to understand
the behavior of the form factors away from $q^2_{max}$.

The paper is  organized  as  follows.  In  section 2 we present a
brief introduction to the light--front relativistic  quark model;
in section 3, we calculate  the  Isgur--Wise  function for $B \ra
D^{(\ast)}$  decays;  in section 4 we study the  heavy--to--light
form  factors in the heavy quark  limit and give  conclusions  in
section 5.

\section{The light--front relativistic quark model}

The light--front  relativistic model \cite{model}
has been recently applied to
many  aspects of heavy  meson  weak  decays  \cite{jaus,ox1,ox2}.
Here we give a brief introduction to the model.

A  ground--state  meson $V(Q{\bar q})$ with spin $J$ in the
light--front quark model can be described by the state vector
\beq
| V (P, J_3, J) \rangle=
\!\!\!\!\!&&\!\!\!\!\!\int d^3\vv{p_1} d^3\vv{p_2}
\ \delta(\vv{P}-\vv{p_1}-\vv{p_2}) \nonumber\\
\!\!\!\!&&\!\!\sum_{\lambda_1,\lambda_2}
\Psi^\an(\vv{P},\vv{p_1},\vv{p_2},\lambda_1,\lambda_2)
|Q(\lambda_1,\vv{p_1} )\ {\bar q}(\lambda_2,\vv{p_2} )\rangle\ .
\label{e21}
\eeq
The quark coordinates are given by
\beq
&&p_{1 +}=x_1 P_+ \ , p_{2 +}=x_2 P_+ \ ,
\ x_1+x_2=1 \ , 0\le x_{1,2}\le 1 \ ,\nonumber\\
&&\vv{p_{1\bot}}=x_1 \vv{P_\bot}+\vv{k_\bot} \ , \
\vv{p_{2\bot}}=x_2 \vv{P_\bot}-\vv{k_\bot} \ .
\label{e22}
\eeq
In the light--front convention,
${\vec P}=(P_+,{\bf P}_\bot)$ where $P_+=P_0+P_z$ and
${\bf P}_\bot=(P_x,P_y)$.
{\sl We denote quark $Q$ by subscript 1 and antiquark  ${\bar q}$
by  2}.  The  quantities   $x_{1,2}$  and  ${\bf  k}^2_\bot$  are
invariant under the kinematic Lorentz  transformations and ${\vec
p_1}+{\vec  p_2}={\vec P}$.  The sum of the  four--momenta of the
quarks is
\beq
M_0^2=(p_1+p_2)^2=
\dv{m_1^2+\vv{k}^2_{\bot}}{x_1}+
\dv{m_2^2+\vv{k}^2_{\bot}}{x_2}\ .
\label{e24}
\eeq
One can introduce a more intuitive quantity, the internal
momentum $\vv{k}=(k_z,\vv{k}_\bot)$
where $k_z$ is defined through
\beq
&&x_1=x=\dv{e_1+k_z}{e_1+e_2} \ \ , \ \
x_2=1-x=\dv{e_2-k_z}{e_1+e_2} \ \ ,\nonumber\\
&&\hspace{1.3cm}e_i=\sqrt{m^2_i+\vv{k}^2} \ \ (i=1,2) \ .
\label{e2n1}
\eeq
Then we have
\beq
M_0=e_1+e_2 \ .
\eeq
The wave function
$\Psi^\an$ is a solution of the relativistic eigenvalue equation
\beq
(e_1+e_2+V_{12})\ \Psi^\an =m_V\ \Psi^\an \ ,
\label{e2n2}
\eeq
where $V_{12}$ is the potential and $m_V$ is the meson mass.

Rotational  invariance  of the wave function for states with spin
$J$ and zero orbital angular momentum  requires the wave function
to have the form \cite{model,jaus} (with $x=x_1$)
\beq
\Psi^{J,J_3}(\vv{P},\vv{p_1},\vv{p_2},\lambda_1,\lambda_2)
=R^\an(\vv{k_\bot},\lambda_1,\lambda_2)
\phi(x, \vv{k_\bot}),
\label{e26}
\eeq
where $\phi(x, \vv{k_\bot})$ is even in $\vv{k_\bot}$ and
\beq
R^{J,J_3}(\vv{k_\bot},\lambda_1,\lambda_2)=\!\!
\sum_{\lambda,\lp}
\langle \lambda_1 | R^\dagger_M(\vv{ k_\bot},m_Q) |\lambda \rangle
\langle \lambda_2 | R^\dagger_M(\vv{-k_\bot},m_{\bar q}) |\lp \rangle
C^{J,J_3}(\dv{1}{2},\lambda;\dv{1}{2},\lambda^\prime).
\label{e27}
\eeq
In Eq.\ (\ref{e27}), \ $C^{J,J_3}(\dv{1}{2},
\lambda;\dv{1}{2},\lambda^\prime)$ is
the Clebsh--Gordan coefficient and the rotation
$R_M(\vv{k_\bot}, m_i)$
on the quark spins is the Melosh rotation \cite{melosh}:
\beq
R_M(\vv{k_\bot},m_i)=
\dv{m_i+x_i M_0-i
\mbox{\boldmath $\sigma$ }\vv{\cdot}(\vv{n\times k_\bot})}
{\sqrt{ (m_i+x_i M_0)^2+\vv{k}^2_{\bot} } } \ ,
\label{e23}
\eeq
where $\vv{n}=(0,0,1)$ and $\boldmath\sigma$
are the Pauli spin matrices. The spin  wave function
$R^{J,J_3}(\vv{k_\bot},\lambda_1,\lambda_2)$
in (\ref{e27}) can also be written as
\beq
R^{J,J_3}(\vv{k_\bot},\lambda_1,\lambda_2)
\!\!\!\!\!&&=
\chi^\dagger_{\lambda_1} R^\dagger_M(\vv{ k_\bot},m_Q)
\ S^\an\ R^{\dagger T}_M(\vv{-k_\bot},m_{\bar q}) \chi_{\lambda_2}
\nonumber\\
\!\!\!\!\!&&=\chi^\dagger_{\lambda_1}\ U^\an_V\ \chi_{\lambda_2} \ ,
\label{e25a}
\eeq
where $S^\an$ is defined by
\beq
S^{J,J_3}=\sum_{\lambda,\lambda^\prime}
|\lambda\rangle \langle \lambda^\prime | \
C^{J,J_3}(\dv{1}{2},\lambda;\dv{1}{2},\lambda^\prime) \ .
\label{e25b}
\eeq
For the pseudoscalar and vector mesons, the non--relativistic
spin matrix is
\beq
S^{0,0}=\dv{i \sigma_2}{\sqrt2} \ \ , \ \
S^{1,\pm 1}=\dv{1\pm\sigma_3}{2}\ \ , \ \
S^{1,0}=\dv{\sigma_1}{\sqrt2} \ .
\label{e27b}
\eeq
Without the Melosh rotation, the spin of the wave function
will be just $S^{J,J_3}$.
The explicit expressions of $U^{1,J_3}$ in Eq.\ (\ref{e25a}) can
be found in ref. \cite{ox1}.

The matrix element of the $B$ meson ($(b{\bar q})$)
decaying to a meson $V(Q{\bar q})$ is
\beq
\!\!\!\!\langle   V(p_V,J_3)|\bar{Q}
\Gamma b | B(p_B) \rangle \!\!\!\!
&&\!\!\!\!=\int dx\ d^2\vv{k_\bot}
\sum_{\lambda_1,\lambda_2}
\dv{ \Psi^{\ast 1,J_3}_V \ \bar u_Q \Gamma u_b\ \Psi^{0,0}_B }
{x}
\nonumber\\
\!\!\!\!&&\!\!\!\!=\int dx\ d^2\vv{k_\bot}
\dv{ \phi^\ast_V \phi_B }{x}
Tr\left[ U^{\dagger 1,J_3}_V\ U_\Gamma\ U^{0,0}_B \right]\ ,
\label{e28}
\eeq
where $U_\Gamma$ is defined by
\beq
{\bar u}^i_Q\ \Gamma\ u^j_b=
\chi^\dagger_i\ U_\Gamma\ \chi_j \ .
\label{e29}
\eeq
In Eq.\ (\ref{e28}), we choose
$q_+=(p_B-p_V)_+ =0$ so $q^2=-{\bf q}^2_\bot$.

Eq.\ (\ref{e28}) is in fact expected to be valid only for ``good"
currents  such as  $\Gamma=\gamma_+,\gamma_+\gamma_5,  $  \ldots.
There  are  contributions  other  than  the  one  given  in  Eq.\
(\ref{e28}) if a current is not a ``good" current \cite{jaus}.

We define the form factors of the
$B(b{\bar q}) \ra P(Q{\bar q})$ transition
between two pseudo--scalar mesons by
\beq
\langle
P(p_P) \!\!\mid {\bar Q} \gamma_\mu b
\mid \!\!B(p_B) \rangle
\!=\!F_1(q^2) (p_B\!+\!p_P\!-\!\dv{(m_B^2-m_P^2)}{q^2} q)_\mu \!+\!
F_0(q^2) \dv{(m_B^2-m_P^2)}{q^2} q_\mu ,
\label{e211}
\eeq
where $q=p_B-p_P$ and one has $F_0(0)=F_1(0)$.
For the $B$ meson transition to a vector meson $V(Q{\bar q})$
we define
\beq
\langle V(p_V,\epsilon)|\bar{Q}
i\sigma_{\mu\nu}q^{\nu}b_{R}|B(p_{B})\rangle\hspace{-0.6cm}
&&=f_{1}(q^{2})i\varepsilon_{\mu\nu\lambda\sigma}
\epsilon^{\ast\nu}p^{\lambda}_{B} p_V^{\sigma}\nonumber\\
&&+ \left[(m^{2}_{B}-m^{2}_{V})\epsilon^{\ast}_{\mu}-
(\epsilon^{\ast}\cdot q)
(p_{B}+p_V)_{\mu}\right]f_{2}(q^{2})\nonumber\\
&&+(\epsilon^{\ast}\cdot q)\!
\left[q_{\mu}-\frac{q^{2}\ (p_{B}+p_V)_{\mu}}{(m^{2}_{B}
-m^{2}_{V})}\right]f_{3}(q^{2}) \nonumber\\
\label{eb1}
\eeq
and
\begin{eqnarray}
&&\hspace{-1cm}\langle
V(p_V,\epsilon) \!\!\mid {\bar Q} \gamma_\mu (1-\gamma_5) b
\mid \!\!B(p_B) \rangle
=\!\!\dv{2 V(q^2)}{m_B+m_V} i \epsilon_{\mu\nu\alpha\beta}
\epsilon^{\ast\nu} p_V^\alpha p_B^\beta -\!\! 2m_V
\frac{(\epsilon^\ast \cdot p_B) }{q^2} q^\mu A_0(q^2)\nonumber\\
&&\hspace{-1cm}-\!\left[ (m_B+m_V) \epsilon^{\ast\mu} A_1(q^2)\!-
\frac{(\epsilon^\ast \cdot p_B) }{m_B+m_V}(p_B+p_V)^\mu A_2(q^2)\!
\!-\!
2m_V \frac{(\epsilon^\ast \cdot p_B)}
{q^2} q^\mu A_3(q^2)\right]\!\!.
\label{eb2}
\end{eqnarray}
Note that the form factor  $f_3(q^2)$  does not contribute to the
decay $B \ra K^\ast  \gamma$.  In Eqs. (\ref{eb1}) and (\ref{eb2})
$\epsilon$ is
the polarization vector of the vector meson, and
\beq
A_3(q^2)=\dv{m_B+m_V}{2 m_V} A_1(q^2)-\dv{m_B-m_V}{2 m_V} A_2(q^2)
\ , \ A_3(0)=A_0(0)\ .
\label{eb3}
\eeq

Using the ``good" current $\Gamma=\gamma_+$ and Eq.\ (\ref{e28}),
we can  obtain  the  following  expression  for the  form  factor
$F_1(0)$:

\beq
F_1(0)=\int dx\ d^2\vv{k_\bot}
\phi^\ast_P(x,\vv{k_\bot}) \phi_B(x,\vv{k_\bot})
\dv{ ({\cal A}_Q {\cal A}_b +{\bf k}^2_\bot) }
{\sqrt{({\cal A}_Q^2\!+\!{\bf k}^2_\bot)(
{\cal A}_b^2\!+\!{\bf k}^2_\bot)}} \ ,
\label{e212}
\eeq
where
\beq
{\cal A}_Q=x\ m_{\bar q}+(1-x) m_Q\ ,
\ {\cal A}_b=x\ m_{\bar q}+(1-x) m_b\ .
\label{e213}
\eeq
The form factor  $F_1(0)$  can be  rewritten  using the  internal
momentum   $\vv{k}=(k_z,   \vv{k}_\bot)$   defined  through  Eq.\
(\ref{e2n1}) instead of the variables $(x,\vv{k}_\bot)$ as
\beq
F_1(0)=
\int d^3\vv{k} \eta^\ast_P(\vv{k}^\prime) \eta_B(\vv{k})
\sqrt{\dv{e_Q e_2^\prime M^b_0}{e_b e_2 M^Q_0}}
\dv{ ({\cal A}_Q {\cal A}_b +{\bf k}^2_\bot) }
{\sqrt{(A^2_Q\!+\!{\bf k}^2_\bot)(A^2_b\!+\!{\bf k}^2_\bot)}}\ ,
\label{e214}
\eeq
where the momentum $\vv{k}$ in the $B$ meson and $\vv{k}^\prime$
in the $P$ meson have the following relation
\beq
\vv{k}_\bot=\vv{k}_\bot^\prime \ \ , \ \
\dv{e_b+k_z}{e_b+e_2}=x=
\dv{e_Q+k_z^\prime}{e_Q+e_2^\prime}\ ,
\label{e215}
\eeq
and for each meson, respectively,
\beq
\phi(x,\vv{k_\bot})=\sqrt{\dv{d\vv{k}_z}{dx}} \eta({\bf k})
\ \ , \ \ \dv{d\vv{k}_z}{dx}=\dv{e_1 e_2}{x_1 x_2 M_0}\ .
\label{e216}
\eeq
In (\ref{e214})  $M^b_0$ and $M^Q_0$  represent  respectively the
quantity  $M_0$  of  (\ref{e24})  for  mesons  $B$ and  $P$.  The
formulae  for the other form  factors can be  obtained  similarly
using the ``good" currents
$\Gamma=\gamma_+ \gamma_5, \gamma_+$ and Eq.\ (\ref{e28}):
\beq
A_0(0)=
\int d^3\vv{k}\ \eta^\ast_V(\vv{k}^\prime)
\hspace{-0.6cm}&&\eta_B(\vv{k})
\ \sqrt{\dv{e_Q e_2^\prime M^b_0}{e_b e_2 M^Q_0}}
\nonumber\\
&&\hspace{-1cm}\dv{{\cal A}_Q
{\cal A}_b\!+\!(2x-1){\bf k}^2_\bot+\!
\dv{2 (m_b+m_Q) (1-x){\bf k}^2_\bot}{W_V} }
{\sqrt{(A^2_Q\!+\!{\bf k}^2_\bot)(A^2_b\!+\!{\bf k}^2_\bot)}}\ ,
\label{e217a}\\
f_1(0)=
\int d^3\vv{k} \ \eta^\ast_V(\vv{k}^\prime)
\hspace{-0.6cm}&&\eta_B(\vv{k})
\ \sqrt{\dv{e_Q e_2^\prime M^b_0}{e_b e_2 M^Q_0}} \nonumber\\
&&\hspace{-1cm}
\dv{ ({\cal A}_Q {\cal A}_b\!+\! x\ {\bf k}^2_\bot+\!
\dv{ (m_b+m_Q) (1-x){\bf k}^2_\bot}{W_V} )}
{\sqrt{({\cal A}_Q^2\!+\!{\bf k}^2_\bot)
({\cal A}_b^2\!+\!{\bf k}^2_\bot)}}
\ ,
\label{e217b}\\
V(0)=
\int d^3\vv{k} \ \eta^\ast_V(\vv{k}^\prime)
\hspace{-0.6cm}&&\eta_B(\vv{k})
\ \sqrt{\dv{e_Q e_2^\prime M^b_0}{e_b e_2 M^Q_0}}
\nonumber\\
&&\hspace{-1cm}
\dv{(m_B\!+\!m_V)(1\!-\!x)({\cal A}_b\!+
\!\dv{{\bf k}^2_\bot}{W_V}\!
+(1-x)(m_b-m_Q) {\bf k}^2_\bot \Theta_V)}
{\sqrt{({\cal A}_Q^2\!+\!{\bf k}^2_\bot)
({\cal A}_b^2\!+\!{\bf k}^2_\bot)}}
\ ,
\label{e217c}\\
A_2(0)=-\int d^3\vv{k} \ \eta^\ast_V(\vv{k}^\prime)
\hspace{-0.6cm}&&\eta_B(\vv{k})
\ \sqrt{\dv{e_Q e_2^\prime M^b_0}{e_b e_2 M^Q_0}}
\ (m_B\!+\!m_V)(1\!-\!x)
\label{e217d} \\
&&\hspace{-4.2cm}
\dv{
(1\!\!-\!\!2 x) {\cal A}_b\!+\!x_2
[m_Q\!+\!(1\!\!-\!\!2 x) m_b\!+\!2 x m_2 ]
\Theta_V \vv{k}_\bot^2
\!\!+\!\!\dv{2
\left( ({\cal B}_Q {\cal A}_b +\vv{k}_\bot^2)
(1+\Theta_V \vv{k}_\bot^2)
+\dv{1}{2}\vv{k}_\bot^2 \right)}{W_V}}
{\sqrt{({\cal A}_Q^2\!+\!{\bf k}^2_\bot)
({\cal A}_b^2\!+\!{\bf k}^2_\bot)}},
\nonumber
\eeq
where
\beq
&&W_V=M^Q_0+m_Q+m_2\ ,\ {\cal B}_Q=x\ m_2 - (1-x) m_Q\ ,
\nonumber\\
&&
\Theta_V=(\dv{d{\tilde \phi}_V}{d{\bf k}^2_\bot})/{\tilde \phi}_V
\ ,\ {\tilde \phi}_V=
\dv{\phi_V}{\sqrt{{\cal A}_Q^2+{\bf k}^2_\bot}}\ .
\label{e218}
\eeq
Note that one can get $A_1(0)$  from $A_0(0)$ and $A_2(0)$  using
Eq.\ (\ref{eb3}) and $f_1(0)=2 f_2(0)$.  In \cite{jaus},  similar
formulae have also been given.

\section{The wave functions}

The meson wave function  $\phi(x,\vv{k_\bot})$ is model dependent
and difficult to obtain; often simple forms are assumed for them.

One reasonable assumption is a Gaussian type of wave function
\beq
\phi(x,\vv{k_\bot})=\eta({\bf k}) \sqrt{\dv{d\vv{k}_z}{dx}}
\ \ , \ \ \eta({\bf k})= \dv{1}{(\pi \omega^2)^{3/4}}
exp\left(-\dv{\vv{k}^2}{2 \omega^2}  \right) \ .
\label{e31}
\eeq
The parameter  $\omega$ is a scale parameter and should be of the
order of $\Lambda_{\rm  QCD}$.  This wave function have been used
in many previous  applications  of the  light--front  quark model
\cite{model,jaus}.  The result are generally quite successful.

A similar wave function is
\beq
\phi(x,\vv{k_\bot})=\eta({\bf k}) \sqrt{\dv{d\vv{k}_z}{dx}}
\ \ , \ \
\eta({\bf k})=N \ exp(-\dv{M^2_0}{2 \omega^2}) \ .
\label{e32}
\eeq
Here $N$ is the normalization constant.  Eq.\ (\ref{e32}) in fact
differs  from Eq.\  (\ref{e31})  only when the two  quarks of the
meson  have  different  masses.  If they  have  equal  mass  i.e.
$m_1=m_2=m$   as  is  the  case  for  the   $\pi$   and   $\rho$,
$M_0^2=(e_1+e_2)^2=4\ e_1^2=4\ e_2^2=4\ (m^2+\vv{k}^2)$.  The two
wave  functions  are  equivalent  since  they  differ  only  by a
constant  factor.  The wave  function  (\ref{e32})  has been also
applied for heavy mesons for which the two quarks  certainly have
different masses \cite{huang,hwang}.

Another possibility is the wave function adopted in \cite{bsw}
\beq
\phi(x,\vv{k_\bot})=
N \sqrt{x(1-x)}
exp\left( -\ \dv{M^2}{2w^2}
\left[ x-\dv{1}{2}-\dv{m_Q^2-m_{\bar q}^2}{2 M^2}\right]^2\right)
\dv{ exp\left(-\ \dv{ \vv{k}^2_{\bot} }{2w^2}\right)}
{\sqrt{\pi w^2} } \ ,
\label{e33}
\eeq
where $M$ is the mass of the meson.

We normalize the wave function to 1:
\beq
1=\int \ji |\phi(x,\vv{k_\bot})|^2
=\int d^3\vv{k} |\eta(\vv{k})|^2.
\label{e34}
\eeq
The wave function $\phi(x,\vv{k_\bot})$
should also satisfy \cite{jaus}
\beq
f_M\hspace{-0.5cm}&&=2 \sqrt{\dv{3}{(2\pi)^3}}
\int dx d^2\vv{k}_{\bot}\ \phi(x,\vv{k_\bot})
\ \dv{{\cal A}_Q}{\sqrt{{\cal A}_Q^2+\vv{k}^2_{\bot}}}
\nonumber\\
\hspace{-0.5cm}&&=2 \sqrt{\dv{3}{(2\pi)^3}}
\int d^3({\bf k})\ \eta(\vv{k})
\ \sqrt{\dv{x_1 x_2 M_0}{e_1 e_2}}
\dv{{\cal A}_Q}{\sqrt{{\cal A}_Q^2+\vv{k}^2_{\bot}}}\ ,
\label{e35}
\eeq
where $f_M$ is the meson decay constant.  When the decay constant
is known,  this  condition  is  usually  one of  several  ways to
determine  the values of  parameters  in the wave  function.  For
heavy mesons such as the $B$ meson it imposes a constraint on the
wave function because the decay constant has the scaling behavior
\cite{heavy}
\beq
f_B \propto \dv{1}{\sqrt{m_b}} \ \ , \ \ m_b \ra \infty \ .
\label{e36}
\eeq
We  have  not  included  perturbative  corrections in
(\ref{e36}).

We now  consider the general  behavior of the  heavy--meson  wave
function  $\phi(x,\vv{k_\bot})$ in the heavy quark limit $m_b \ra
\infty$.  Here  we  take  the  $B$  meson  as  an   example.  The
distribution amplitude $\int d^2\vv{k}_\bot  \phi(x,\vv{k}_\bot)$
of a heavy meson is known to have a peak near $x\simeq 1$.  If we
denote the $x$ coordinate of the peak in $\phi(x,\vv{k_\bot})$ by
$x_0$ and the width of the peak in $x$ by  $\Delta_x$,  then $x_0
\ra 1$ and  $\Delta_x  \ra 0$ as $m_b \ra \infty$.  Thus the wave
function behaves like a $\delta$--function  in $x$.  On the other
hand,   $\phi(x,\vv{k}_\bot)$   vanishes  if   $\vv{k}^2_\bot  >>
\Lambda^2_{\rm  QCD}$.  To see this in some detail  consider Eq.\
(\ref{e2n1}).  If we use $(k_z, \vv{k}_\bot)$ as the coordinates,
then we expect  that  $|\vv{k}|$  should, in  general,  be of the
order of  $\Lambda_{QCD}$.  Thus from  Eq.\  (\ref{e2n1}),  it is
easy  to  see   that   the  $x$   coordinate   of  the   peak  of
$\phi(x,\vv{k_\bot})$ behaves like
\beq
x=1- \dv{\tilde \Lambda}{m_b}+\cdots
\label{e37}
\eeq
where ${\tilde \Lambda}$ is a function of $\vv{k}_\bot^2$  but it
is of order of $\Lambda_{QCD}$.

All three wave  functions  listed  before have the above  general
feature.  However, as we show in the appendix, the wave functions
(\ref{e31})  and (\ref{e33})  satisfy the scaling law (\ref{e36})
but  (\ref{e32})  does  not.  The wave  function  (\ref{e33})  is
obtained  assuming  factorization  with  respect  to the spin and
orbital  motion  \cite{nr,nrsx}  which  is not  the  case  in the
light--front relativistic quark model.  Thus we will only use the
wave function (\ref{e31}) in the following sections.

\section{The Isgur--Wise function of $B \ra D^{(\ast)}$ decays}

With the  formalism and the wave  function  given in the previous
sections, it is now straightforward to calculate the form factors
of the $B \ra D^{(\ast)}$ semileptonic decays.
We study the behavior of these
form factors in the limit $m_b \ra \infty$ and $m_c \ra  \infty$.

We   use   the   wave    function    (\ref{e31})    and    define
$r=\dv{m_c}{m_b}$.  Also, we take  $m_B=m_b$,  $m_D=m_c$, and set
the scale parameters for the heavy mesons equal,  $\omega_H\equiv
\omega_B=\omega_D$.  (The same $\omega_D$  holds for both $D$ and
$D^\ast$    mesons).   In   the   integrals    (\ref{e214})   and
(\ref{e217a})--(\ref{e217d}),  $|\vv{k}|$ for the $B$ meson is of
the order of the scale  parameter  $\omega_H$ and
$k_z^\prime$ for the $D^{(\ast)}$ is given by
\beq  k_z^\prime=
\dv{1}{2}   \left(r (k_z-\sqrt{m_2^2   +\vv{k}^2_\bot+k_z^2}\ )
+\dv{(k_z+\sqrt{m_2^2+\vv{k}^2_\bot+k_z^2}\ )}{r}     \right)
+O(\dv{1}{m_b}) \ .
\label{e41}
\eeq
(Note that $\vv{k}_\bot^\prime=\vv{k}_\bot$.)
In terms of the variable  $x$ the  integrand  in these  integrals
peaks at
\beq
x=1-\dv{\sqrt{m_2^2+\vv{k}^2_\bot+k_z^2}-k_z}{m_b}+\cdots  \  .
\label{e42}
\eeq
and  $M^c_0$  and  $M^b_0$  in these  integrals
become
\beq M^b_0 \ra m_b \ \ , \ \ M^c_0 \ra m_c=r\ m_b\ , \ \ e_b
\ra m_b \ \ , \ \ e_c \ra m_c=r\  m_b \ \ .
\label{e43}
\eeq
In   (\ref{e217a})--(\ref{e217d})   all  terms   proportional  to
$\dv{1}{W_V}$ can be neglected in the heavy quark limit.  Thus,
\beq
&&\hspace{-0.8cm}
F_1^{B  D}(0)=A_0^{B D^\ast}(0)=f_1^{B D^\ast}(0)= \int d^3\vv{k}
\eta^\ast_D(\vv{k}^\prime) \eta_B(\vv{k})
\sqrt{\dv{e_2^\prime}{e_2}}   \nonumber\\
&&\hspace{+5cm}
\dv{({\cal  A}_c {\cal A}_b+{\bf k}^2_\bot)}{\sqrt{({\cal
A}_c^2\!+\!{\bf k}^2_\bot)({\cal A}_b^2\!+\!{\bf k}^2_\bot)}}\ ,
\label{e44a}\\
&&\hspace{-0.8cm}A_2^{B D^\ast}(0)=V^{B
D^\ast}(0)=\int d^3\vv{k} \
\eta^\ast_D(\vv{k}^\prime)\eta_B(\vv{k}) \
\sqrt{\dv{e_2^\prime}{e_2}}
\nonumber\\
&&\hspace{+3.2cm}\dv{(m_B\!+\!m_{D^\ast})(1\!-\!x)({\cal  A}_b\!+
m_b (1-r) (1-x) {\bf  k}^2_\bot  \Theta_{D^\ast})}  {\sqrt{({\cal
A}_c^2\!+\!{\bf k}^2_\bot)({\cal  A}_b^2\!+\!{\bf  k}^2_\bot)}}
\ .
\label{e44b}
\eeq

We are left  with two sets of form  factors.  Now we need to show
that these two sets of form factors are equal, as required by the
heavy quark  symmetry.  For the special  mass ratio $r=1$, we can
show  analytically  that  all the  above  form  factors  equal 1.
However,  it is not  easy  to  show  that  the  form  factors  in
(\ref{e44a})  and  (\ref{e44b})  are equal for an arbitrary ratio
$r$.  Nevertheless we can use numerical  calculation to show that
these form  factors  are  indeed  equal \cite{num}.
Thus  there is only one
independent form factor in the model, as required by the
heavy quark symmetry.
This form factor  does not depend on the heavy  quark
masses $m_b$ and $m_c$ but their ratio $r$:
\beq
h\hspace{-0.5cm}&&={\tilde h}(r, m_2,\omega_H)\nonumber\\
&&=F_1^{B  D}(0)=A_0^{B  D^\ast}(0)= A_1^{B
D^\ast}(0)= A_2^{B D^\ast}(0)=V^{B  D^\ast}(0)=f_1^{B D^\ast}(0)
\nonumber\\
&&=\int  d^3\vv{k}   \eta^\ast_D(\vv{k}^\prime)
\eta_B(\vv{k})\sqrt{\dv{e_2^\prime}{e_2}}\ \dv{({\cal  A}_c
{\cal A}_b+{\bf  k}^2_\bot)}{\sqrt{({\cal A}_c^2\!+\!{\bf
k}^2_\bot)({\cal A}_b^2\!+\!{\bf k}^2_\bot)}}\ .
\label{e45}
\eeq

In  \cite{nr,nrsx}  it is argued that the  knowledge  of the form
factor $h$ at  $q^2=0$  suffices  to  determine  the  Isgur--Wise
function  in the whole  kinematic  region.  The basic  idea is as
follows.  The Isgur--Wise  function  depends only on the velocity
product $v \cdot v^\prime$
\beq
v \cdot v^\prime =\dv{m_B^2+m_D^2-q^2}{2 m_B m_D}\ \ .
\label{e46}
\eeq
We define
\beq
\ \ y=v \cdot v^\prime (q^2=0)=\dv{r^2+1}{2 r}\ .
\label{e47}
\eeq
Keeping  $q^2=0$ fixed, $y$ changes as the mass ratio $r$ changes
in the interval  $[1,\infty]$ which can cover the whole kinematic
region in weak decay.  Writing $h$ in terms of $y$
\beq
h=h(y, m_2, \omega_H)
\label{e48}
\eeq
gives the Isgur--Wise function to be
\beq
\xi(y)=\sqrt{\dv{2}{y+1}} \ h(y, m_2, \omega_H)\ .
\label{e49}
\eeq
Hence we obtain the  Isgur--Wise  function  $\xi(y)$ in the whole
kinematic  region even though in the light--front  quark model we
can only calculate the form factors at $q^2=0$.

It is easy to see the  Isgur--Wise  function  obtained  this  way
satisfies  $\xi(1)=1$  since when  $r=1$  then  $h=1$ and  $y=1$.
Also, the Isgur--Wise  function satisfies Bjorken's constraint on
the   derivative    $\rho^2=-\xi^\prime(1)=1/4-h^\prime(1,   m_2,
\omega_H)>1/4$  because the overlap of normalized  wave functions
can  not  be   larger   than   one,   i.e
$h^\prime(1,m_2,\omega_H)<0$ \cite{nr,nrsx}.

The final Isgur--Wise function is calculated numerically.  We use
the light quark masses  obtained by fitting  $f_\pi$ and $f_\rho$
\cite{jaus}:  $m_2=0.25$GeV.  For the heavy  quark  masses,  $f_D
\simeq   200$MeV   for   $m_c=1.6$GeV   and   $\omega_D=0.45$GeV.
Similarly,   $f_B   \simeq   190$MeV   for    $m_b=4.8$GeV    and
$\omega_B=0.55$GeV.  These  values of $f_B$ and $f_D$  lie in the
ranges of recent  lattice  results  \cite{lattice2,lattice3}.  We
use the  above  masses  for $b$ and $c$  quarks  and will  choose
$\omega_H=\omega_D=\omega_B=0.55$GeV  to work in the heavy  quark
limit.  The  Isgur--Wise  function is calculated  in the range $1
\le   y   \le    \dv{r_0^2+1}{2    r_0}   \simeq    1.67$   where
$r_0=\dv{m_c}{m_b}=0.33$.  To   see   the   dependence   of   the
Isgur--Wise  function on the light quark masses, we also give the
result  corresponding to  $m_2=0.30$GeV.  In Figure 1 we show the
Isgur--Wise   function   $\xi(y)$  for  the  $B  \ra  D^{(\ast)}$
semileptonic  decays.  For  comparison we also show the functions
$(\dv{2}{y+1})^m$ $(m=1, 2)$ which correspond to the single-- and
double--pole--like form factors in the heavy quark limit.

The $\rho^2$ of the Isgur--Wise function is
\beq
\rho^2=-\xi^\prime(1)=1.23\ ,
\ \ m_2=0.25{\rm GeV}\ ,
\nonumber\\
\rho^2=-\xi^\prime(1)=1.27\ ,
\ \ m_2=0.30{\rm GeV}\ .
\eeq

For   comparison,  we  list  in  Table  I  a  number  of  recent
calculations    of    $\rho^2$.    The    lattice    calculations
\cite{lattice}   and  most  of  the  relativistic   quark  models
\cite{rela}  tend to give larger values  ($\rho^2 \gapx 1$).  The
values  from  QCD  sum  rule  calculations   \cite{sumrule}   are
generally  somewhat smaller.  The  non--relativistic  quark model
\cite{isgw}  gives  $\rho^2  \simeq 0.67$ (with the large  recoil
effect used to fit the $\pi$ electromagnetic charge radius) but a
relativistic  modification of this model carried out by Close and
Wambach gives $\rho^2 \simeq 1.19$ \cite{rela}.

\section{Form factors of heavy--to--light decays
in the heavy quark limit}

In this  section we study the form  factors  of  heavy--to--light
transitions $B \ra P(Q{\bar q})$ and $B \ra V(Q{\bar  q})$, where
$m_b \ra  \infty$  as  before,
but the  quark $Q$ is a light quark.

Consider  the  form  factors  of   equations   (\ref{e214})   and
(\ref{e217a}--\ref{e217d}).  Since the wave  function  of the $B$
meson  peaks  near $x  \simeq  1$, i.e.  $x \ra 1$ when  $m_b \ra
\infty$, the {\sl integrands}  usually also have a peak.  The $x$
coordinate of this peak takes, in general, the form
\beq
x=1-O(\dv{1}{m_b^n})\ \ , \ \ n>0\ ,
\label{e51}
\eeq
where $n$  depends on the  specific  form of the wave  functions.
(With the wave  function we used the integrand
for the heavy--to--heavy  decays has $n=1$.)
Due to (\ref{e51}) the quantity $M_0^{Q}$
\beq
M_0^{Q} \propto m_b^{\frac{\scriptstyle n}{\scriptstyle 2}}
\ra \infty \ .
\label{e52}
\eeq
Therefore, regardless of the specific form of the wave functions,
terms in the  expressions  for the form factors  proportional  to
$\dv{1}{W_V}$  can be  ignored  when  $m_b \ra  \infty$.  This is
similar to what we saw in the last  section for the heavy--to--heavy
transitions.  For  heavy--to--light  transitions the form factors
in (\ref{e217a})--(\ref{e217d}) now reduce to
\beq
F_1(0)
&=&\int d^3\vv{k} \eta^\ast_P(\vv{k}^\prime) \eta_B(\vv{k})
\sqrt{\dv{e_Q e_2^\prime M^b_0}{e_b e_2 M^Q_0}}
\dv{ ({\cal A}_Q {\cal A}_b +{\bf k}^2_\bot) }
{\sqrt{({\cal A}_Q^2\!+\!{\bf k}^2_\bot)
({\cal A}_b^2\!+\!{\bf k}^2_\bot)}}\ ,
\label{e53a}\\
A_0(0)=f_1(0)
&=&\int d^3\vv{k} \eta^\ast_V(\vv{k}^\prime) \eta_B(\vv{k})
\sqrt{\dv{e_Q e_2^\prime M^b_0}{e_b e_2 M^Q_0}}
\dv{ ({\cal A}_Q {\cal A}_b +{\bf k}^2_\bot) }
{\sqrt{({\cal A}_Q^2\!+\!{\bf k}^2_\bot)
({\cal A}_b^2\!+\!{\bf k}^2_\bot)}}\ ,
\label{e53b}\\
A_2(0)=V(0)
&=&\int d^3\vv{k} \ \eta^\ast_V(\vv{k}^\prime)\eta_B(\vv{k})
\ \sqrt{\dv{e_Q e_2^\prime M^b_0}{e_b e_2 M^Q_0}}
\nonumber\\
&&\hspace{+0cm}\dv{(m_B\!+\!m_V)(1\!-\!x)({\cal A}_b\!+
(1-x) (m_b-m_Q) {\bf k}^2_\bot \Theta_V)}
{\sqrt{({\cal A}_Q^2\!+\!{\bf k}^2_\bot)
({\cal A}_b^2\!+\!{\bf k}^2_\bot)}}\ ,
\label{e53c}
\eeq
in the limit $m_b \ra  \infty$.  Thus in the  light--front  quark
model we  obtain  two  independent  form  factors  for  $B\ra  V$
transitions regardless of the specific form of the wave function.
Because of Eq.  (\ref{e53b}) we found  \cite{ot1,ox1,xu} that the
ratio  $\cal I$ which  relates  the decay  rate of $B \ra  K^\ast
\gamma$  and that of $B \ra \rho e {\bar  \nu_e}$ at $q^2=0$ is 1
in the SU(3) flavor symmetry limit.

In general there should be four  independent form factors for the
heavy--to--light $B\ra V$ transition and, in particular, the four
form factors $A_1(q^2)$,  $A_2(q^2)$, $V(q^2)$ and $f_1(q^2)$ are
independent of each other  \cite{ot2}.  Here, in the light--front
relativistic  quark  model we obtain  only two  independent  form
factors  due  to  the   vanishing   of  terms   proportional   to
$\dv{1}{W_V}$.  This can be traced back to the  treatment  of the
quark  spins  (\ref{e26}--\ref{e25a}),  which  corresponds  to  a
weak--binding limit \cite{korner}.  This is a known approximation
in  the  light--front  quark  model  \cite{sjc}.  In  the  matrix
element  for  the  heavy--to--light  transition  $B  \ra  V$  the
integrand  has  contributions  only from  $x_1=x \ra 1$ as $m_b \ra
\infty$.  Thus the Melosh rotation (\ref{e23}) for the transition
quarks $b$ and $Q$ becomes
\beq
R_M(\vv{k}_\bot, m_b) \ra 1
\ ,\ \ R_M(\vv{k}_\bot, m_Q) \ra 1 \ ,
\label{e513}
\eeq
even  though  the quark $Q$ is light.  Thus the  Melosh  rotation
affects only the spectator  quarks.

We now use the wave  function  (\ref{e31})  to  obtain  the  form
factors,  the  details  of which are given in the  appendix.  The
final  expression  for the form  factors in the $m_b \ra  \infty$
limit is
\beq
&&\hspace{-1.2cm}f_1(0)=V(0)=A_0(0)=A_2(0) (=A_1(0))=\nonumber\\
&&\hspace{-1.2cm}\dv{4 \cdot 2^{11/12}}{\sqrt3}\ r_V^{-7/12}
\ (\dv{m_2}{m_b})^{2/3}
\ exp \left[ \dv{-3\ (2 r_V)^{1/3} m_2^{4/3} m_b^{2/3}
+(3+4\ r_V\!)\ m_2^2+ 2\ m_Q^2}
{16\ \omega_V^2} \right]\!,
\label{e544}
\eeq
where   $r_V=\omega_V^2   /  \omega_B^2$   with   $\omega_V$  and
$\omega_B$  being the meson  scale  parameters  in wave  function
(\ref{e31}).  The  replacement  of  $\omega_V$ by $\omega_P$  and
$r_V$ by  $r_P=\omega_P^2 / \omega_B^2$ gives $F_1(0)  (=F_0(0)$.
It is  interesting  to note  that  though there are in general
two sets of form factors for $B\ra V$ transition,
these two sets of form factors become equal when we use the wave
function (\ref{e31}).
We attribute this equality to the specific form of the
wave  function (\ref{e31}).  We see no reason that
this is generally  true for  arbitrary  wave  functions.
Obviously, all the
form factors have the following dependence on $m_b$:
\beq
\dv{exp(-m_b^{2/3}  a)}{m_b^{2/3}} \label{e545}
\eeq
where $a=\dv{3\  (2  r_L)^{1/3}   m_2^{4/3}}{16\   \omega_L^2}$
($L=V  \  {\rm  or}\  P$).

The dependence of the  heavy--to--light  form factors on $m_b$ is
interesting  because it allows us to compare with the  prediction
from the pole--dominance assumption for the heavy--to--light form
factors.  For  example,  Burdman  and  Donoghue   \cite{bd}  have
pointed out the inconsistency between the scaling behavior of the
heavy--to--light    form   factor   at   $q^2_{max}$    and   the
single--pole--like form factors used in the BSW model \cite{bsw}.
One knows \cite{iw}
\beq
&&F_0(q^2_{max}) \propto m_b^{-1/2}
\ ,\ A_1(q^2_{max}) \propto m_b^{-1/2}
\ ,\ \nonumber\\
&&F_1(q^2_{max}) \propto m_b^{1/2} \ ,
\  V(q^2_{max}) \propto m_b^{1/2}
\ ,\ A_2(q^2_{max}) \propto m_b^{1/2}
\ ,
\label{e512}
\eeq
but  all  form  factors  in the BSW  model  behave like $\propto
m_b^{-1}$  at  $q^2=0$   when  $m_b  \ra   \infty$.
Thus  in  BSW  model  the
single--pole--like $q^2$ dependence can certainly not produce the
scaling law (\ref{e512}).

The light--front  quark model is a relativistic quark model which
contains  many  important  ingredients  not  included  in the BSW
model.  Obviously, a single--pole--like $q^2$ dependence combined
with  (\ref{e512})  can not produce our results.  In our opinion,
this is an indication that a single--pole--like  $q^2$ dependence
may not be  correct,  at  least  in  region  far  away  from  the
zero-recoil   point   $q^2_{max}$,  like  $q^2=0$.  In  fact  the
pole--dominance  assumption  is generally  expected to be correct
only near the zero recoil region.  The actual $q^2$ dependence of
the form  factors  far away from  $q^2_{max}$  could be much more
complicated.
When compared with (\ref{e512})
our  result  indicates  that  the  set  $F_1(q^2)$,
$V(q^2)$  and   $A_2(q^2)$   may  have  similar   type  of  $q^2$
dependence,  while the set $F_0(q^2)$ and  $A_1(q^2)$  also has a
similar $q^2$  dependence, but different from the first set.  For
example, $F_1(q^2)$ may increase as $q^2$ faster than $F_0(q^2)$.
Similarly,  $V(q^2)$  and  $A_2(q^2)$  may  increase  faster than
$A_1(q^2)$.

The form factors in the  heavy--to--light  $B \ra P$ and $B\ra V$
semileptonic decays and their dependence on the $b$ quark mass in
the $m_b \ra \infty$ limit have been studied by many other people
\cite{lattice3,xu2,nardulli,cz,rueckel,ball,orsay}.            In
particular, in  \cite{cz,rueckel}  the  light--cone  QCD sum rule
gives $F_1(0) \propto m_b^{-3/2}$, which is not in agreement with
a single pole  extrapolation.  However, in  \cite{nardulli} a QCD
sum  rule   calculation   gives  $F_1(0)   \propto   m_b^{-1/2}$,
consistent with a single pole assumption for the whole  kinematic
region.  Though there are  differences  in all these  studies, it
seems  most  people  do agree  that  form  factors  $V(q^2)$  and
$A_2(q^2)$  may increase  faster than the form factor  $A_1(q^2)$
\cite{nardulli,ball,orsay} as $q^2$ increases.

Finally,  we give the  numerical  results of the form  factors at
$q^2=0$ for the transitions $B\ra \pi$,  $B\ra\rho$, $B\ra K$ and
$B\ra K^\ast$:
\beq
&&
F^{B \ra \pi}_1(0)=0.26\ ,
\ \ f_1^{B \ra \rho}(0)=0.28\ ,
\ \ V^{B \ra \rho}(0)=0.32\ , \nonumber\\
&&
A^{B \ra \rho}_0(0)=0.30\ ,
\ \ A^{B \ra \rho}_1(0)=0.21\ ,
\ \ A^{B \ra \rho}_2(0)=0.18\ ;
\eeq
\beq
&&
F^{B\ra K}_1(0)=0.34\ ,
\ \ f_1^{B \ra K^\ast}(0)=0.37\ ,
\ \ V^{B \ra K^\ast}(0)=0.42\ , \nonumber\\
&&
A^{B \ra K^\ast}_0(0)=0.40\ ,
\ \ A^{B \ra K^\ast}_1(0)=0.29\ ,
\ \ A^{B \ra K^\ast}_2(0)=0.24\ .
\eeq
For the $B$ meson we have used $m_b=4.8$GeV, $\omega_B=0.55GeV$;
for $\pi$, $\rho$, $K$ and $K^\ast$, the parameters are taken
from \cite{jaus}: $m_u=m_d=0.25$GeV,
$\omega_\pi=\omega_\rho=0.32$GeV, $m_s=0.37$GeV,
$\omega_K=\omega_{K^\ast}=0.39$GeV.

\section{Conclusion}

In  this  paper  we  have   studied  the  form   factors  of  the
heavy--to--heavy  and  heavy--to--light  weak  transitions in the
light--front  relativistic quark model.  For the heavy--to--heavy
$B\ra D^{(\ast)}$ transitions we have shown that the form factors
satisfy the heavy quark symmetry  relations.  We have  calculated
the  corresponding   Isgur--Wise   function.  The  slope  of  the
Isgur--Wise  function  agrees  with most other  calculations.  We
have  also  studied  the  heavy--to--light  $B\ra P$ and $B\ra V$
transitions.  For the transition  $B\ra V$ the model  produces at
most two  independent  form  factors.  In general  there are four
independent  form factors;  this  reduction  comes from using the
weak  binding  limit and is  independent  of the  choice  of wave
function.  With a specific  wave  function,  we have  derived the
dependence of the form factors (at $q^2=0$) on the $b$ quark mass
in the  $m_b  \ra  \infty$  limit.  This  dependence  can  not be
produced  by  extrapolating  the  scaling  behavior  of the  form
factors at $q^2_{max}$  using the single--pole  assumption.  This
shows that the $q^2$  dependence  of the form  factors in regions
far away  from the  zero--recoil  could be much more  complicated
than that predicted by the single--pole assumption.
When compared with the scaling behavior of the form factors
at $q^2_{max}$ our result suggests, for example, that
$F_1(q^2)$ increases as $q^2$ faster than $F_0(q^2)$ and
similarly,  $V(q^2)$  and  $A_2(q^2)$ increase  faster than
$A_1(q^2)$.

\vspace{.3in}
\centerline{ {\bf Acknowledgement}}
This work was in part supported by the Natural Sciences and
Engineering Council of Canada.

\newpage

\begin{center}
{\bf Appendix }
\end{center}

{\bf 1. Wave functions and the scaling law (\ref{e36})}

Here we examine the wave functions (\ref{e31}--\ref{e33})
to see if they satisfy the scaling law
(\ref{e36}).

We first look at the wave function (\ref{e31}).
Fot the $B$ meson the wave function $\eta_B(\vv{k})$
in the integral (\ref{e35}) does not lead to any $m_b$ dependence.
Since in the heavy quark limit,
\beq
x_1=x \ra 1\ ,
\ M_{0B} \ra m_b\ ,\ e_1 \ra m_b,
\label{e38}
\eeq
and
\beq
x_2=1-x \propto m_b^{-1}\ ,
e_2\propto  m_b^{0} \ \ , \ \ {\cal A}_b \propto m_b^{0}\ ,
\label{e39}
\eeq
we can easily obtain the scaling law (\ref{e36}). Note that here
$\propto m_b^{0}$ means independent of $m_b$.

Next, we look at the wave function (\ref{e32}).
In fact we can show that wave functions of the following type
for the $B$ meson
do not satisfy the scaling law (\ref{e36}):
\beq
\phi_B(x,t)=N_B\ g(x,t)\ exp(-f(x,t)), \ \ t=\vv{k}_\bot^2\ ,
\label{e310}
\eeq
where
\beq
f(x,t)=\dv{(M^{b}_0)^2}{2 \omega_B^2}
\label{e311}
\eeq
and $g(x,t)$ is some polynomial, rational or irrational function.
We assume the function $\phi_B(x,t)$
peaks when $f(x,t)$ has a minimum. Suppose
$f(x,t)$ is at its minimum when $x=x_0$ and $t=t_0$, then
\beq
x_0=1-\dv{m_2}{m_b+m_2}\ \ , \ \ t_0=0\ .
\label{e312}
\eeq
We expand $f(x,t)$ around $x=x_0$ and $t=t_0$:
\beq
f(x,t)=f(x_0,t_0)+\dv{1}{2}\dv{d^2}{dx^2}\!f(x_0, t_0) (x-x_0)^2
+\dv{d}{dt}\!f(x_0, t_0)\ t+\cdots
\label{e313}
\eeq
where
\beq
f(x_0, t_0)=\dv{(m_b+m_2)^2}{2 \omega_B^2}
,\dv{d^2}{dx^2}\!f(x_0, t_0)
= \dv{(m_2+m_b)^4}{m_2 m_b\omega_B^2}
,\dv{d}{dt}\!f(x_0, t_0)
=\dv{(m_2+m_b)^2}{2 m_2 m_b\omega_B^2}.
\label{e313b}
\eeq
We can show higher order terms in $x$ can be neglected.
(Obviously there is no higher order term in $t$.)
The derivatives $\dv{d^2}{dx^2}\!f(x_0, t_0)$ and
$\dv{d}{dt}\!f(x_0, t_0)$ determine the width of the
wave function $\phi(x,t)$ at $(x_0,t_0)$.
The wave function can be considered as zero when
$(x-x_0)$ is of order larger than $m_b^{-3/2}$ and $t-t_0$
of order larger than $m_b^{-1}$. This means the width of the
wave function (\ref{e32}) becomes zero in both $x$ and $t$
when $m_b \ra \infty$.

Now we can determine $N_B$ of (\ref{e310}) from the
normalization condition (\ref{e34}).
We introduce $\delta$ and $\lambda$ to remove the dependence
of the integration variables on $m_b$:
\beq
x=x_0+(\dv{m_2}{m_b})^{3/2}\ \delta \ \ \ ,
t=t_0+(\dv{m_2}{m_b})\ m_2^2\ \lambda \ .
\label{e314}
\eeq
With the new variables $(\delta, \lambda)$,
the exponent $f(x,t)$ is now
\beq
f(x,t)=\dv{(m_2+m_b)^2}{2 \omega_B^2}
+\dv{m_2^2}{2 \omega_B^2}(\delta^2+\lambda)+\cdots
\label{e317}
\eeq
and the wave function (\ref{e310})  becomes
\beq
\phi_B(x,t)
&&\hspace{-0.6cm}=g(x,t) exp(-f(x,t)) \nonumber\\
&&\hspace{-0.6cm}\ra m_b^j\ {\bar g}(\delta,\lambda)\
exp\left[ -\dv{(m_2+m_b)^2}{2 \omega_B^2}
-\dv{m_2^2}{2 \omega_B^2}(\delta^2+\lambda) \right]\!,
\label{e318}
\eeq
where ${\bar g}(\delta,\lambda)$ has no dependence on $m_b$
and $j$ is some number depending on the specific
form of $g(x,t)$. For the wave function (\ref{e32}) which
is an example of (\ref{e310})
\beq
\phi_B(x,t)&&\hspace{-0.6cm}=
N_B \sqrt{\dv{d\vv{k}_z}{dx}}\ exp(-f(x,t))\nonumber\\
&&\hspace{-0.6cm}\ra
N_B\ m_b\ exp\left[ -\dv{(m_2+m_b)^2}{2 \omega_B^2}
-\dv{m_2^2}{2 \omega_B^2}(\delta^2+\lambda) \right]\!.
\label{e319}
\eeq
Note
\beq
x \ra 1\ , x_2 \ra \dv{m_2}{m_b}\ , \ M^b_0 \ra m_b\ ,
\ e_b \ra m_b\ , \ e_2 \ra m_2\ .
\label{e320}
\eeq
With the new variables $(\delta,\lambda)$
the normalization condition (\ref{e34}) becomes
\beq
1=\int \ji |\phi_B(x,\vv{k_\bot})|^2
=\pi\ m_2^2\ (\dv{m_2}{m_b})^{5/2} \int d\delta\ d\lambda
|\phi_B(x,\vv{k_\bot})|^2 \ .
\label{e315}
\eeq
and (\ref{e35}) becomes
\beq
f_B\hspace{-0.6cm}&&=2 \sqrt{\dv{3}{(2\pi)^3}}
\int dx\ d^2\vv{k}_{\bot}
\phi_B(x,\vv{k_\bot})
\dv{{\cal A}_b}{\sqrt{{\cal A}_b^2+\vv{k}^2_{\bot}}}
\nonumber\\
&&=2 \sqrt{\dv{3}{(2\pi)^3}}\ \pi\ m_2^2\ (\dv{m_2}{m_b})^{5/2}
\int d\delta\ d\lambda
\ \phi_B(x,\vv{k_\bot})\ .
\label{e316}
\eeq
Note the factor $\dv{{\cal A}_b}{ \sqrt{{\cal A}_b^2+\vv{k}^2_\bot} }
\ra 1$ in
the $m_b \ra \infty$ limit.
In both (\ref{e315}) and (\ref{e316})
the factor $\pi\ m_2^2\ (\dv{m_2}{m_b})^{5/2}$
comes from the integration--variable transformation.
{}From equations (\ref{e318}) and (\ref{e315}--\ref{e316})
we get
\beq
f_B \propto m_b^{-5/4}
\label{e321}
\eeq
for the wave function of type (\ref{e310}).

We can show that the wave function (\ref{e33})
satisfies the scaling law (\ref{e36}).

\vskip 10pt
{\bf 2. Heavy to light form factors in the heavy quark limit}
\vskip 10pt

Now we use the wave function (\ref{e31}) to
study the dependence of the heavy--to--light form factors
on $m_b$.
The integrands in (\ref{e53a})--(\ref{e53c})
are all of the form
${\tilde g}(k_z,t) exp(-{\tilde f}(k_z,t)$ with
\beq
{\tilde f}(k_z,t)={\tilde f_B}(k_z,t)+{\tilde f_L}(k_z,t)
\label{e54p}
\eeq
where
\beq
&&{\tilde f_B}(k_z,t)=
\dv{\vv{k}^2}{2 \omega_B^2}=\dv{r_L\ \vv{k}^2}{2 \omega_L^2}\ ,
\ \ \ {\tilde f_L}(k_z,t)=\dv{\vv{k}^{\prime\ 2}}{2 \omega_L^2}\ ,
\nonumber\\
&&{\bf k}^2=k_z^2+t\ ,
\ \ \ {\bf k}^{\prime 2}=k_z^{\prime 2}+t\ ,
\ \ \ t={\bf k}_\bot^2\ ,
\ \ \ r_L=\dv{\omega_L^2}{\omega_B^2}\ ,
\label{e54}\\
&&\hspace{-1cm}
(L=P \ {\rm for}\ B \ra P\ {\rm transition},
 L=V \ {\rm for}\ B \ra V\ {\rm transition})
\nonumber
\eeq
and ${\tilde g}(k_z,t)$ can be
determined from (\ref{e214}) and (\ref{e217a})--(\ref{e217d}).
The internal momentum ${k^\prime_z}$ can be expressed in terms of
$k_z$ and $t$ through (\ref{e215}).

The minimum point of ${\tilde f}(k_z,t)$
is where the integrand peaks.
Suppose the coordinates of the minimum point of ${\tilde f}(k_z,t)$ is
$(k_{z 0},t_0)$.
One can then expand ${\tilde f}(k_z,t)$ near
$(k_{z 0},t_0)$
\beq
{\tilde f}(k_z,t)
={\tilde f}(k_{z 0},t_0)+
\dv{1}{2}
\dv{d^2}{dk_z^2}{\tilde f}(k_{z 0},t_0)
(k_z-k_{z0})^2
+\dv{d}{d t}{\tilde f}(k_{z 0},t_0)
\ (t-t_0)+\cdots
\label{e55}
\eeq
One can show higher order terms in (\ref{e55}) can be neglected.
We find
\beq
t_0=0 \ \ , \ \
k_{z0}=-2^{-4/3} r_L^{-1/3}
m_2^{2/3} m_b^{1/3} \ .
\label{e56}
\eeq
The coefficients in the expansion (\ref{e55}) are
\beq
{\tilde f}(k_{z 0},t_0)=(\dv{1}{2 \omega_L^2})
\left[ \dv{3}{8} (2 r_L)^{1/3} m_2^{4/3} m_b^{2/3}
-(\dv{3+4 r_L}{8}) m_2^2-\dv{1}{4} m_Q^2 \right]
,\nonumber\\
\dv{d^2}{d k_z^2}{\tilde f}(k_{z 0},t_0)=\dv{3}{\omega_B^2}
\ \ , \ \ \dv{d}{d t}{\tilde f}(k_{z 0},t_0)=
\dv{1}{2^{8/3} \omega_L^{4/3} \omega_B^{2/3}} (\dv{m_b}{m_2})^{2/3}\ .
\label{e58}
\eeq
Thus the width at the peak
is independent of $m_b$ in $k_z$ but
is of order $m_b^{-2/3} (\ra 0)$ in $t$.
Hence in the $m_b \ra \infty$ limit,
the exponential function $exp(-{\tilde f}(k_z,t))$ behaves like
a $\delta$ function in $t$ and the contribution to the integrands
in (\ref{e53a})--(\ref{e53c}) comes only from $t_0=0$.
In terms of the coordinate $x$ the peak is at
\beq
x_0=1-(2\ r_L)^{-1/3} (\dv{m_2}{m_b})^{2/3}+\cdots\ ,
\ x_{2 0}=1-x_0=(2\ r_L)^{-1/3} (\dv{m_2}{m_b})^{2/3}+\cdots\ .
\label{e57}
\eeq
It is interesting to notice that
though $x_0 \ra 1$ as in the heavy--to--heavy decays,
$x_{2 0}=1-x_{0}$ has a different dependence on $m_b$.
It is not of order $m_b^{-1}$ but $m_b^{-2/3}$. Again, in terms
of $x$, one can show that the contribution to the integrand comes
only from $x=x_0$.

Because of (\ref{e56}) and (\ref{e57}), the integrands in
(\ref{e53a})--(\ref{e53c}) become much simpler.
To obtain an analytical expressions for
integrals in (\ref{e53a})--(\ref{e53c})
we again introduce a couple of new variables $(\xi,\lambda)$
to get rid of the $m_b$ dependence in $(k_z,t)$
\beq
k_z=k_{z0}+m_2\ \xi \ \ ,
\ \ t=t_0+(\dv{m_2}{m_b})^{2/3}\ m_2^2\ \lambda\ .
\label{e59}
\eeq
In terms of the new variables, ${\tilde f}(k_z,t)$ now becomes
\beq
{\tilde f}(k_z,t)=
{\tilde f}(k_{z 0},t_0)
+(\dv{3\ r_L}{2}\cdot\dv{m_2^2}{\omega_L^2} )\delta^2
+(\dv{ (2 r_L)^{1/3} }{8}\cdot\dv{m_2^2}{\omega_L^2}) \lambda
+\cdots
\label{e510}
\eeq
We substitute (\ref{e59}) into the integrals
(\ref{e53a})--(\ref{e53c}) and
keep the leading terms in the expansion in $m_b$.
The integrals (\ref{e53a})--(\ref{e53c}) then become
\beq
&&\hspace{+0.8cm}F_1(0)=\!\!
\int\! d\delta\ d\lambda
\ exp\left(
-{\tilde f}(k_{z 0},t_0)
-(\dv{3\ r_P}{2}\cdot\dv{m_2^2}{\omega_P^2} )\delta^2
-(\dv{ (2 r_P)^{1/3} }{8}\cdot\dv{m_2^2}{\omega_P^2}) \lambda
\right)\!C,
\nonumber\\
&&\hspace{-0.7cm}A_0(0)=f_1(0)=\!\!
\int\! d\delta\ d\lambda\ exp\left(
-{\tilde f}(k_{z 0},t_0)
-(\dv{3\ r_V}{2}\cdot\dv{m_2^2}{\omega_V^2} )\delta^2
-(\dv{ (2 r_V)^{1/3} }{8}\cdot\dv{m_2^2}{\omega_V^2}) \lambda
\right)\!C,
\nonumber\\
&&\hspace{-0.7cm}A_2(0)=V(0)=\!\!
\int\! d\delta\ d\lambda\ exp\left(
-{\tilde f}(k_{z 0},t_0)
-(\dv{3\ r_V}{2}\cdot\dv{m_2^2}{\omega_V^2} )\delta^2
-(\dv{ (2 r_V)^{1/3} }{8}\cdot\dv{m_2^2}{\omega_V^2}) \lambda
\right)\!C
\nonumber\\
&&\hspace{2.3cm}\left[
(\dv{m_b}{m_2})^{1/3}
\dv{(2^{8/3} \omega_V^2 - r_V^{1/3} m_2^2 \lambda))}
{8 r_V^{1/3} \omega_V^2 }
-
\dv{(2 \omega_V^2 \delta -
2^{-8/3} r_L^{1/3} m_2^2 \lambda)}
{\omega_V^2 } \right]\!,
\eeq
where
\beq
C=\dv{2^{3/4}\ r_L^{1/4}\ m_2^{11/3} }
{2 \sqrt{\pi} m_b^{2/3} \omega_L^3}
\ \ (L=P\ {\rm or}\ V)\ .
\eeq
The final expression
for the form factors in the $m_b \ra \infty$ limit is
\beq
&&\hspace{-1.2cm}F_1(0)=\nonumber\\
&&\hspace{-1.2cm}\dv{4 \cdot 2^{11/12}}{\sqrt3}\ r_P^{-7/12}
\ (\dv{m_2}{m_b})^{2/3}
\ exp \left[ \dv{-3\ (2 r_P)^{1/3} m_2^{4/3} m_b^{2/3}
+ (3+4 r_P)\ m_2^2+ 2\ m_Q^2 }
{16\ \omega_P^2} \right]\!,
\nonumber\\
&&\hspace{-1.2cm}A_0(0)=f_1(0)=A_2(0)=V(0)=\nonumber\\
&&\hspace{-1.2cm}\dv{4 \cdot 2^{11/12}}{\sqrt3}\ r_V^{-7/12}
\ (\dv{m_2}{m_b})^{2/3}
\ exp \left[ \dv{-3\ (2 r_V)^{1/3} m_2^{4/3} m_b^{2/3}
+ (3+4 r_V)\ m_2^2+ 2\ m_Q^2 }
{16\ \omega_V^2} \right]\!.
\label{e511b}
\eeq
The difference between
$F_1(0)$ and the form factors of $B \ra V$ transition
comes from the scale parameter.

Finally, we give a brief explanation about why the two
groups of form factors $A_0(0)=f_1(0)$ and $A_2(0)=V(0)$
are equal in our calculation. Since in terms of the coordinates
$x$ and $t$ the contribution to the integrand comes only from
$x_0$ and $t_0$ given by (\ref{e56}) and (\ref{e57}) we have
\beq
A_b \ra m_b\ x_{2 0} \ , \ A_Q \ra m_2\ ,
\ {\bf k}_\bot^2 \ra 0 \ .
\label{e57b}
\eeq
Thus in the integrals (\ref{e53a})--(\ref{e53c})
\beq
&&\hspace{1cm}\dv{ ({\cal A}_Q {\cal A}_b +{\bf k}^2_\bot) }
{\sqrt{({\cal A}_Q^2\!+\!{\bf k}^2_\bot)
({\cal A}_b^2\!+\!{\bf k}^2_\bot)}} \ra 1\ ,
\label{e61}\\
&&\hspace{-2cm}\dv{(m_B\!+\!m_V)(1\!-\!x)({\cal A}_b\!+
(1-x) (m_b-m_Q) {\bf k}^2_\bot \Theta_V)}
{\sqrt{({\cal A}_Q^2\!+\!{\bf k}^2_\bot)
({\cal A}_b^2\!+\!{\bf k}^2_\bot)}}\nonumber\\
&&\hspace{2cm}\ra\dv{ [ m_b\ x_{2 0}+m_2+
m_b\ x_{2 0}\ t\ \Theta_V ] }{m_2} \ .
\label{e62}
\eeq
In the square brackets of (\ref{e62})
we keep the $m_2$ term because the other two terms,
though of higher order in $m_b$, cancel each other, as we begin to
show now. The definition of $\Theta_V$ is given in (\ref{e218}):
\beq
\Theta_V=(\dv{d{\tilde \phi}_V}{d{\bf k}^2_\bot})/{\tilde \phi}_V
\ ,\ {\tilde \phi}_V=
\dv{\phi_V}{\sqrt{{\cal A}_Q^2+{\bf k}^2_\bot}}
=G_V(k_z,t) exp\left(-{\tilde f_V}(k_z,t)\right) \ ,
\label{e63}
\eeq
where
\beq
G_V(k_z,t)=\dv{\sqrt{\dv{d\vv{k}^\prime_z}{dx}} }
{(\pi \omega^2)^{3/4} \sqrt{{\cal A}_Q^2+{\bf k}^2_\bot}}\ .
\label{e64}
\eeq
Thus
\beq
\Theta_V=-\dv{d}{d t}{\tilde f_V}(k_z,t)+
(\dv{d}{d t}G_V(k_z,t))/G_V(k_z,t)\ .
\label{e65}
\eeq
One can show that the leading contribution to $\Theta$ comes from
$-\dv{d}{d t}{\tilde f_V}(k_z,t)$ at $(k_{z 0},t_0)$, i.e.
\beq
\Theta_V=-\dv{d}{d t}{\tilde f_V}(k_{z 0},t_0)\ .
\label{e66}
\eeq
The coefficient $\dv{d}{d t}{\tilde f}(k_{z 0},t_0)$ in
(\ref{e58}) is the sum of $\dv{d}{d t}{\tilde f_B}(k_{z 0},t_0)$
and $\dv{d}{d t}{\tilde f_V}(k_{z 0},t_0)$
\beq
\dv{d}{d t}{\tilde f}(k_{z 0},t_0)=
 \dv{d}{d t}{\tilde f_B}(k_{z 0},t_0)
+\dv{d}{d t}{\tilde f_V}(k_{z 0},t_0)\ ,
\label{e67}
\eeq
but since
\beq
\dv{d}{d t}{\tilde f_V}(k_{z 0},t_0)=
\dv{1}{2^{8/3} \omega_L^{4/3} \omega_B^{2/3}} (\dv{m_b}{m_2})^{2/3}
>>\dv{d}{d t}{\tilde f_B}(k_{z 0},t_0)=\dv{1}{2 \omega_B^2}\ ,
\label{e68}
\eeq
$\dv{d}{d t}{\tilde f}(k_{z 0},t_0)$ is equal to
$\dv{d}{d t}{\tilde f_V}(k_{z 0},t_0)$
\beq
\dv{d}{d t}{\tilde f}(k_{z 0},t_0) \ra
\dv{d}{d t}{\tilde f_V}(k_{z 0},t_0)=
\dv{1}{2^{8/3} \omega_L^{4/3} \omega_B^{2/3}} (\dv{m_b}{m_2})^{2/3}\ ,
\label{e69}
\eeq
as given in (\ref{e58}).
Thus in the expression (\ref{e53c}),
when being integrated over $t$,
terms proportional to
$m_b x_{2 0}$ and $m_b x_{2 0} t \Theta_V$ of (\ref{e62}) become
\beq
\int d t\ [m_b\ x_{2 0} + m_b\ x_{2 0}\
t\ (-\dv{d}{d t}{\tilde f_V}(k_{z 0},t_0) ]
\ exp \left(-\dv{d}{d t}{\tilde f_V}(k_{z 0},t_0) t \right)=0\ .
\eeq
Hence the terms $m_b x_{2 0}$ and $m_b x_{2 0} t \Theta_V$
in (\ref{e62}) cancel and (\ref{e62}) becomes 1, equal to
(\ref{e61}). This is why the two sets of form factors
$f_1(0)=A_0(0)$ and $V(0)=A_2(0)$ are equal.

\newpage

{\bf Figure:}
The Isgur--Wise function $\xi(y)$ for the
$B \ra D^{(\ast)}$ semileptonic decays.
The dashed and dotted lines correspond to $\xi(y)$ with
$m_2=0.25$GeV and $m_2=0.30$GeV, respectively;
the solid lines represents the single-- and double--pole
functions $\dv{2}{y+1}$ and $(\dv{2}{y+1})^2$.

\begin{picture}(-2000,100)(80,-100)
\includegraphics{/users/qiping/maple/figure.ps}
\end{picture}

\newpage

\baselineskip=25pt

\begin{table}
\baselineskip=25pt

\vspace{1.5cm}

\begin{center} Table I.
The slope of the Isgur--Wise function for
$B \ra D^{(\ast)} e {\bar \nu}_e$ decays.
\end{center}

\begin{center}

\begin{tabular}{|c | c|} \hline\hline
& \\
 This work
& 1.23\ \ ($m_2=250$MeV)\\
& 1.27\ \ ($m_2=300$MeV)\\
& \\
\hline
Ahmady \et \cite{rela}             & 0.54--1.5              \\
Bernard \et \cite{lattice}         & $1.41\pm 0.19\pm 0.41$ \\
Blok and Shifman \cite{sumrule}    & 0.5--0.8               \\
Close and Wambach \cite{rela}      & $ 1.19\pm 0.02$        \\
Holdom and Sutherland\cite{rela}   & 1.24--1.36  	    \\
Huang and Luo \cite{sumrule}       & $ 1.01\pm 0.02$        \\
Isgur \et \cite{isgw}              & 0.64                   \\
Ivanov and Mizutani \cite{rela}    & 0.42--0.82             \\
Jin \et \cite{rela}                & 0.97                   \\
Kiselev \cite{rela}                & 1.25                   \\
Kugo \et \cite{rela}               & 1.8--2.0               \\
Mannel \et \cite{old}              & $1.77\pm 0.74$         \\
Narison \cite{sumrule}             & 0.52--0.92             \\
Neubert \cite{sumrule}             & $0.7\pm 0.2 $          \\
Rosner \cite{old}                  & $1.44\pm 0.41$         \\
Sadzikowski and Zalewski\cite{rela}& 1.24                   \\
UKQCD \cite{sumrule}               & $1.2^{+7}_{-2} $       \\
\hline\hline
\end{tabular}

\end{center}

\end{table}


\begin{thebibliography}{99}
\bibitem{num}
We have tried numerical calcuations with different precision.
The form factors are always equal up to the digit
where these numerical calculations are reliable.
\end{thebibliography}
\end{document}